\documentclass{optica-article}

\journal{opticajournal} 

\articletype{Research Article}

\usepackage{lineno}
\usepackage{gensymb}    

\begin{document}

\title{Structural color filters with compensated angle-dependent shifts}

\author{Katarína Rovenská,\authormark{1,2} Filip Ligmajer,\authormark{1,2,*} Beáta Idesová,\authormark{1,2}, Peter Kepič,\authormark{1,2} Jiří Liška,\authormark{1,2} Jan Chochol,\authormark{3} and Tomáš Šikola\authormark{1,2}}

\address{\authormark{1}Central European Institute of Technology, Brno University of Technology, Purkyňova 123, Brno 621 00, Czech Republic\\
\authormark{2}Institute of Physical Engineering, Faculty of Mechanical Engineering, Brno University of Technology, Technická 2, Brno 612 00, Czech Republic\\
\authormark{3} onsemi, 1. máje 2634, 756 61 Rožnov pod Radhoštěm, Czech Republic}

\email{\authormark{*}filip.ligmajer@vutbr.cz} 


\begin{abstract*} 
Structural color filters use nano-sized elements to selectively transmit incident light, offering a scalable, economical and environmentally friendly alternative to traditional pigment- and dye-based color filters. However, their structural nature makes their optical response prone to spectral shifts whenever the angle of incidence varies. We address this issue by introducing a conformal VO\textsubscript{2} layer onto bare aluminum structural color filters. The insulator-metal transition of VO\textsubscript{2} compensated the spectral shift of the filter’s transmission at a 15° tilt with 80\% efficiency. Unlike solutions that require adjustment of the filter's geometry, this method is versatile and suitable also for existing structural filters. Our findings also establish tunable materials in general as a possible solution for angle-dependent spectral shifts.

\end{abstract*}

\section{Introduction}
Color separation and filtering are essential components of various electronic devices like displays, cameras, or spectrometers. In most of these devices, color filtering is based on various subpixel filter arrays (like Bayer or PenTile patterns), which are typically realized with pigments and dyes. The problem with this approach is that each of the colors in the filter must be processed in an individual set of fabrication steps, which naturally complicates and increases the cost of the fabrication process \cite{gather_2007}. Moreover, the chemicals in the active layer are often toxic, non-recyclable, and suffer from degradation \cite{hao_1997,humans_2010}. A cheaper and more scalable alternative to these filters may be their structural counterparts, in which scattering and absorption of the nano-sized features are tailored precisely to suppress the transmission of the selected wavelengths. \cite{joo_2020,song_2019,daqiqeh_2021} 

Amongst structural color filters, multilayer systems with no lateral structure have been studied widely, as their scattering and absorption properties can be easily varied by the choice of materials and thicknesses of the layers forming them \cite{park_2021,rana_2020,zheng_2019,lee_flexible_2019,mao_2016}. The simple geometry of these filters relates to simple fabrication, which is important especially for devices with large footprints, but it works solely for single-color filters. Should there be more colors present in a color-filtering device, the necessary variance of the layer thicknesses and/or materials would lead to unwanted increase of fabrication complexity. Such a demand can be fulfilled by structural filters relying on nanopatterns (discs, rods, wires, holes, etc.), which allow for straightforward control over spatial distribution of colors in the filter. \cite{li_structural_2021,huo_photorealistic_2020,fleischman_high_2019,wang_full_2017,horie_2017,shen_structural_2015} Such color filters are composed of subpixels with varying structural parameters (nanostructure’s size, shape, or array spacing), which can be engineered to match the desired color pattern of the device while all structures are fabricated in a single CMOS-compatible step. Furthermore, various possibilities for dynamic color tunability have been proposed, utilizing liquid crystals \cite{driencourt_electrically_2020,nasehi_2019}, flexible substrates \cite{tseng_2017,song_2017}, and optically \cite{wang_2016,gao_2018}, electrically \cite{sreekanth_2021,xu_2016} or thermally \cite{ding_2016} tunable platforms. 

Unfortunately, the performance of all structural filters depends on the angle of incident illumination. In contrast to the multilayer systems, where strategies how to overcome this issue have been demonstrated repeatedly \cite{gao_2021,song_2020,lee_2019,yang_2016,yang_compact_2015}, only a few solutions were reported for nanopattern-based filters. Multistep nanowire arrays \cite{motogaito_fabrication_2021,yue_highly_2017,sun_wide-incident-angle_2014} and subwavelength nanocavities \cite{lee_subwavelength_2016,wu_angle-insensitive_2013} were presented, which withheld their color-filtering capabilities up to 60° while paying the price in terms of only a single allowed incoming polarization state. Angle- and polarization-independent optical response at the same time was demonstrated in color filters that excluded the influence of nanostructure periodicity by randomly distributing them \cite{ye_angle-insensitive_2015} and in thin metallic patches on dielectric/metal layers that can only support angle-insensitive localized surface plasmon resonances \cite{yang_angle_2016,yang_design_2014}.

Here, we present structural filters which utilize new means of angle-dependence mitigation: holey aluminum color filters with thin conformal vanadium dioxide (VO\textsubscript{2}) layer that can compensate spectral shifts induced by illumination under non-zero angles of incidence. The angle-dependence compensation is ensured by the phase change in VO\textsubscript{2}, which can be induced by various sorts of external stimuli. We first discuss properties of bare holey aluminum filters and their performance under non-zero angles of incidence. We then demonstrate the effects of the conformally deposited VO\textsubscript{2} layer on the color-filtering properties under normal and non-zero angle conditions and analyze the phase-change-based compensation of the unwanted spectral shifts. Owing to the phase change in VO\textsubscript{2}, we were able to retrieve the original spectral position of the color filter's transmission peak up to 15° while maintaining 80\%  of the transmission achieved at normal incidence. We find the presented solution for angle-dependent spectral shifts highly flexible, as the conformal VO\textsubscript{2} layer can be deposited onto color-filters of arbitrary geometries. Additionally, the presented compensation principle can be easily extended outside the region of visible wavelengths if other phase change materials are considered and it can also be actively controlled by fast external stimuli. 

\section{Methods}
Transmission spectra of the presented structural color filters were numerically simulated using Ansys Lumerical FDTD software. The geometry and material composition of the model directly corresponds to the fabricated samples, as described in the main text. The lateral size of the simulated pixel was 10 \micro m $\times$ 10 \micro m, the overall height of the simulation region was 1.8 \micro m and it was enclosed within boundaries formed by perfectly matched layers on all sides. The dielectric functions of fused silica and aluminum were taken from the built-in database, which is based on Palik's handbook. Vanadium dioxide was modelled using our own ellipsometry data \cite{kepic_optically_2021}. The whole structure was illuminated by a Gaussian-type source (under scalar approximation) with a waist of 6 \micro m diameter, under the appropriate angle of incidence. The transmission of the filter was calculated from a power monitor located 400 nm below the aluminum layer, still inside the fused silica substrate.   

For the fabrication of switchable structural color filters, we first evaporated a 100 nm thick aluminum layer on a fused silica substrate. Using electron beam lithography, we created a polymer mask (CSAR AR-P 6200.09, Allresist) which served for selective reactive ion etching (RIE) of the underlying aluminum layer. In RIE, BCl\textsubscript{3}/Cl\textsubscript{2} gas mixture was let in for 5~s (30/5~sccm, pressure 6~mTorr) to etch through the native oxide layer and then we etched the aluminum layer for 22~s (30/10~sccm, pressure 4~mTorr). After resist removal, we used thermal atomic layer deposition from a pre-heated TDMAV (tetrakis(dimethylamino)vanadium) precursor to conformally cover the structural color filters with a 30~nm thin VO\textsubscript{2} layer. Immediately after deposition, the sample was annealed in a tube furnace at 500\textdegree C for 15 minutes with the inlet of 15~sccm of O\textsubscript{2} to ensure high quality switching of optical properties in VO\textsubscript{2} at its phase change.

To measure the optical transmission with variable angles of light incidence, we used a custom-built optical setup with a tiltable sample holder. The inevitable lateral and axial shifts during sample tilting were compensated by a fine mechanical positioning system. A 100 W halogen lamp and Andor Shamrock spectrometer were used for illumination and analysis of the light coming from a (5 $\times$ 12)~\textmu m\textsuperscript{2} area of the color filter selected by a set of apertures. For the sample heating, we added an annular resistive heating element covered by thermoconductive tape onto our sample holder. The sample temperature was controlled in a feedback loop by adjustments of the electric current in the device and measured by a small thermal probe attached to the heating element. 

\section{Bare holey aluminum color filters}

\begin{figure}[b]
\centering\includegraphics[width=\linewidth]{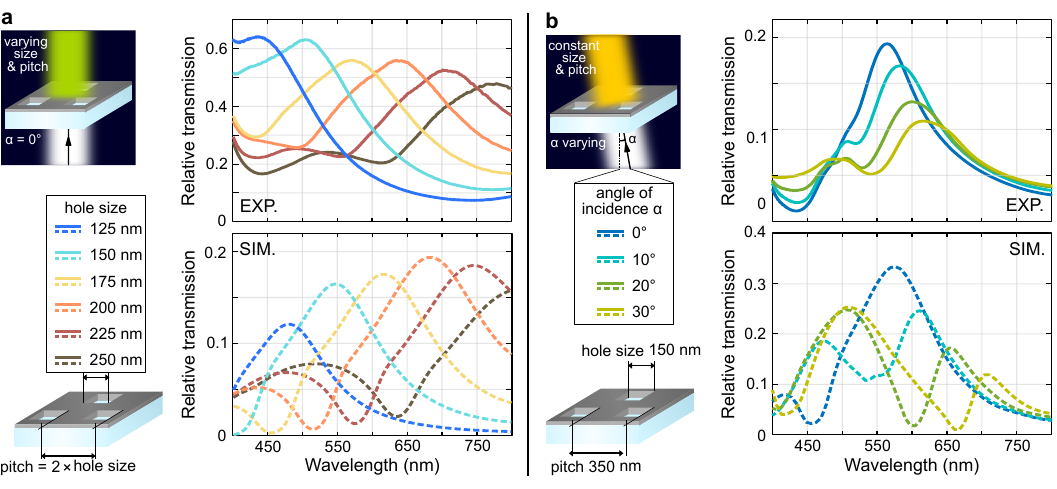}
\caption{Performance of bare holey aluminum color filters in experiment (solid lines) and in numerical simulations (dashed lines). a) Upon normal incidence, the position of the transmission peak maximum can be finely tuned via modifications of size and pitch of the holes in the aluminum film. The larger the hole size and pitch of the holes, the more red-shifted spectra can be observed. b) For a holey aluminum color filter with fixed structural parameters, its transmission maximum decreases and shifts towards red as the angle of incidence increases.}
\label{fig1}
\end{figure}

Holey metallic films are known to function as structural color filters owing to the plasmonic resonances excited in them \cite{ebbesen_extraordinary_1998}. Fine modifications of the size, shape or lateral arrangement of the holes in the metallic film enable us to tune the resonant transmission wavelength of such color filters with high precision \cite{yokogawa_plasmonic_2012}. However, these adjustments can come at a cost: For example, an increase of the hole array’s pitch (in search of a spectral red-shift) leads, inevitably, to a loss of signal as less and less holes are present in the film. Similarly, modification of holes’ size cause uneven transmission distribution across individual colors as the smaller holes transmit less light. In our color filter design, we chose to keep a fixed set of structural parameters ($\mathrm{pitch = 2 \times hole~size}$) in order to maintain a constant fill-factor and thus achieve minimal changes in the intensity of the transmission peaks. To assure the high transmission of the filters, a square shape of the nanohole was preferred over a circular one. In addition, we arranged the nanosized holes in a hexagonal lattice, which has higher fill-factor and thus higher transmission in comparison to a standard square lattice. Favorably, hexagonal lattices also suppress the polarization sensitivity due to their higher degree of symmetry. 

Apart from the pattern geometry, the used materials have also a significant impact on the color filtering. In our case, we chose to work with aluminum as it is CMOS-compatible, cheap, and overall highly accessible. Additionally, aluminum supports strong plasmonic resonances in a broad spectral range from ultraviolet to visible \cite{knight_2014}. In all our experiments, we maintained the thickness of the aluminum film t\textsubscript{Al} = 100 nm as a compromise between possible transmission through the unpatterned film (in case the film is too thin) and fabrication difficulties for laterally small holes (in case the film is too thick).

In Fig.~\ref{fig1}a, we demonstrate the color filtering abilities of the holey structural filters described above. In the simplest case of normal light incidence ($\alpha=0$\textdegree), filters with variable hole size can evenly transmit a wide range of colors across the visible part of the electromagnetic spectrum. In agreement with the known theory and the results obtained from numerical simulations, the measured relative transmission spectra experience red-shift as the hole spacing and hole size increase. Note that the vertical axes in the shown graphs represent relative transmission of the filters in relation to the empty setup with no sample. 

In practical applications, the illumination of a color-filtering element is often not restricted only to normal incidence. The light coming under non-zero angle $\alpha$ encounters slightly different sample geometry in comparison to the normal illumination. As a result, the peaks in transmission spectra tend to red-shift as the angle of incidence increases. To explore the angular dependence of our color-filtering system, we tilted the sample (otherwise positioned normal to the beam) by an angle $\alpha$ while keeping the remaining elements of the optical setup intact. As we gradually increase the angle up to $30$\textdegree, the relative transmission peak of the color filter indeed experiences a red-shift, as shown in Fig.~\ref{fig1}b. Favorably, the red-shifted peaks tend to stay close to the transmission curve of $\alpha=0$\textdegree~ and are thus not introducing new colors into the transmission spectra. Inevitably, the transmission peaks get smaller as less light is transmitted through the tilted systems (light encounters holes that are effectively smaller). These observations are again in line with the results of numerical simulations (see Methods).

\section{VO\textsubscript{2}-filled color filters}

\begin{figure}[b]
  \centering\includegraphics[width=\linewidth]{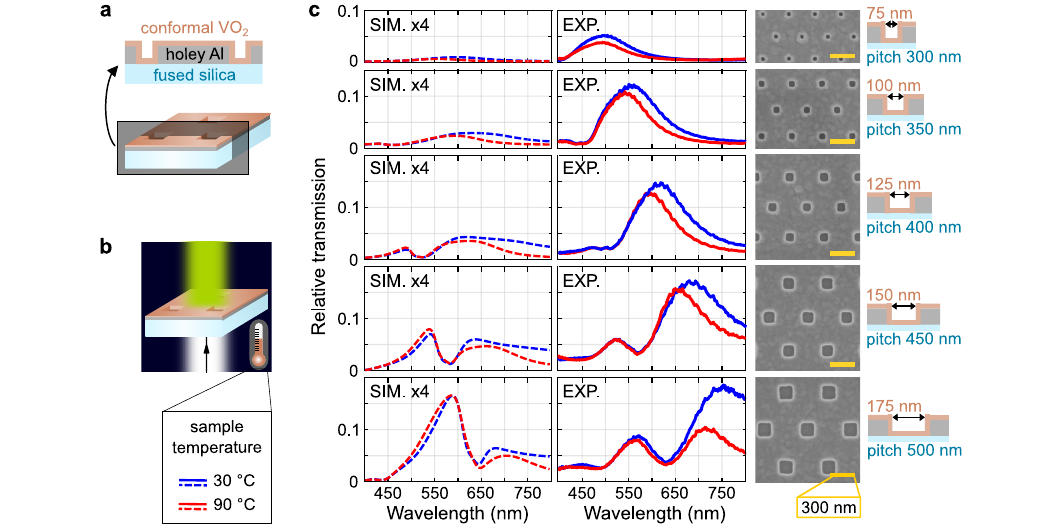}
  \caption{Holey aluminum filters with conformal VO\textsubscript{2} infill. a) Cross-section of the sample geometry. b) Visualisation of the studied situation. c) The phase change of VO\textsubscript{2} allows for two distinguishable responses from a single-color pixel under the normal light incidence. When heated past the insulator-metal transition temperature to 90 \textdegree C, a steady blue-shift can be observed in filters designed for activity across the whole visible spectrum. Simulated results are in the left column, experimental results are in the middle column. SEM micrographs of the corresponding filters are also shown on the right. All scale bars are 300 nm.}
  \label{fig2}
\end{figure} 

As discussed above, the angle-dependent spectral shifts of structural color filters are likely to introduce major flaws when applied in modern electronic devices. Although structural color filters with various remedies for this matter have been proposed, they all rely on a rigid selection of structural parameters that avoid angular-dependent shifts at the cost of overall efficiency. We propose a new type of remedy for this issue in the form of materials with tunable optical response. In this approach, we compensate the angle-dependent spectral shifts using a phase-change material layer that is conformally deposited onto a color filter independently from its previous fabrication, as suggested in Fig.~\ref{fig2}a). A new degree of freedom thus arises, as the externally triggered phase change of the deposited material is associated with the change of its refractive index, and this indirectly compensates the unwanted angle-dependent spectral shifts. Because these shifts lean towards red, as seen above, we need a material which allows for shift of the transmission peak towards blue in order to compensate for them. Conveniently, this property is inherently exhibited by VO\textsubscript{2} when it undergoes its insulator to metal transition around 67~\textdegree C \cite{kepic_optically_2021}. Note, the phase transition in VO\textsubscript{2} is easily reversible and repeatable with minimal degradation. \cite{Liu_2013} Moreover, it does not have to be triggered only by external heating and fast electrical or optical pulses can be also used. \cite{Haglund_2021, atwater_2019}

\begin{figure}[b]
  \centering\includegraphics[width=\linewidth]{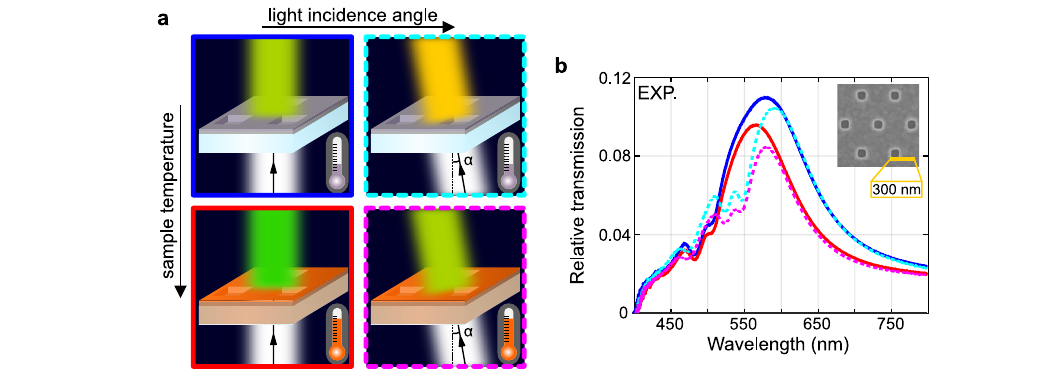}
  \caption{VO\textsubscript{2}-filled structural color filter can be used for compensation of spectral red-shifts caused by non-normal sample illumination. a) A schematic of the experimental concept. b) Relative transmission of the VO\textsubscript{2}-filled color filter corresponding to the four settings shown in a) – curve colors correspond to the frame colors in a). The VO\textsubscript{2}-filled color filter illuminated at a non-zero angle of incidence regains its original transmission peak position when heated. A SEM micrograph of the measured VO\textsubscript{2}-filled color filter is shown in the inset.}
  \label{fig3}
\end{figure}

Naturally, as one partially swaps the air in the original filter design for the VO\textsubscript{2}, which has relatively higher index of refraction, the color-filter's spectral performance at default state shifts towards longer wavelengths. To ensure that the filters remain active for the previously designed colors even after the addition of a $n>1$ layer, bare Al/air filters must be fabricated with smaller holes and/or narrower pitches to begin with, so that the deposition of VO\textsubscript{2} red-shifts the spectral response to the desired color-filtering performance. We preferred to keep the designed hole sizes unchanged and focused on reducing the pitch between the holes of the filter. Taking into account that VO\textsubscript{2} is absorptive in the visible, the higher fill-factor of the color-filters is only beneficial for their overall transmission. We numerically simulated the spectral response of the newly designed VO\textsubscript{2}-filled filters under normal incidence of light for a temperature below and above the insulator-metal transition of VO\textsubscript{2} as illustrated in Fig.~\ref{fig2}b) and the desired blue-shift stemming from the phase transition was indeed observed (see the left column of plots in Fig.~\ref{fig2}c). We therefore decided to also fabricate the structures (see Methods) and measure their color-filtering abilities experimentally under normal incidence of light. As shown in the right column of plots in Fig.~\ref{fig2}c, the measurements yielded better results than predicted by simulations and we observed the expected blue-shift caused by heat-induced phase change of the VO\textsubscript{2}. To each line of plots in Fig.~\ref{fig2}c, a respective scanning electron microscope image of the conformally overdeposited color filter's is presented, with a scale bar of 300 nm length.

Finally, we quantified the sought-after phase-change compensation of the angle-dependent shifts in an experiment. By the simultaneous application of heat at a non-zero angle of incident illumination, the corresponding blue-shift and red-shift, respectively, cancel each other and hence the same color should be obtained, as schematically illustrated in Fig.~\ref{fig3}a. To confirm this hypothesis, we repeated the angular dependence experiment with the VO\textsubscript{2}-filled color filters in two settings: below and above the phase transition temperature of VO\textsubscript{2}. In Fig.~\ref{fig3}b, the solid blue curve shows the transmission with normal illumination and below the phase transition temperature of VO\textsubscript{2}. The dashed cyan curve in Fig.~\ref{fig3}b represents the unwanted red-shifted transmission at $\alpha=15$°. However, by switching the VO\textsubscript{2} to metallic state (dashed magenta curve in Fig.~\ref{fig3}b), the maximum of the color filter’s relative transmission falls back to its default position at $\alpha=0$°, albeit at the expense of the slightly lowered overall transmission.

\section{Conclusion}
In conclusion, we demonstrated facile compensation of unwanted angle-dependent spectral shifts in holey aluminum structural color filters by means of conformal VO\textsubscript{2} layer deposited over the bare aluminum filters. By driving VO\textsubscript{2} through its insulator-metal transition we were able to retrieve the spectral position corresponding to normal light incidence even when the light was coming at an angle of $15$°. The newly regained transmission peak has approximately twice as narrow FWHM than the original one while the relative transmission is reduced by 20 \%. The advantage of this approach lies in its compatibility with production process of traditional structural filters, simply as an additional fabrication step that can be applied to any structural color filter disregard of its geometry. Additionally, the conformal layer can be added to the color filter selectively using masks during its fabrication and thus, one could benefit from the simultaneous existence of both compensated and original optical transmission of the tilted color filter. Overall, we find the flexibility of this method for dealing with angular dependence very high and in future outlook, we envision the use of other phase change materials with wider tunability options that could support the compensation of angle-dependent shifts to the greater extent.


\begin{backmatter}
\bmsection{Funding}
This work has been supported by the Grant Agency of the Czech Republic (21-29468S).

\bmsection{Acknowledgments}
CzechNanoLab project LM2023051 funded by MEYS CR is gratefully acknowledged for the financial support of the measurements/sample fabrication at CEITEC Nano Research Infrastructure.

\bmsection{Disclosures}
 The authors declare no conflicts of interest.

\bmsection{Data availability} Data underlying the results presented in this paper are available in ...





\end{backmatter}


\bibliography{Al_clanek_lib}

\end{document}